\documentclass[preprint,prb,groupedaddress,nofootinbib,showpacs,rotate]{revtex4}

\usepackage{amssymb} 
\usepackage{amsmath} 

\usepackage[dvips]{graphicx}
\begin{document}

\title{ Relativistic energy of a moving spherical capacitor}

\author{Nelson R. F. Braga}
\email{braga@if.ufrj.br}
\affiliation{Instituto de F\'{\i}sica,
Universidade Federal do Rio de Janeiro, Caixa Postal 68528, RJ 21941-972 -- Brazil}
\author{Gustavo Sophia }
\email{gustavosophia@gmail.com}
\affiliation{Instituto de F\'{\i}sica,
Universidade Federal do Rio de Janeiro, Caixa Postal 68528, RJ 21941-972 -- Brazil}


\begin{abstract} 
We discuss the relativistic transformation of the energy of a charged spherical capacitor.
The energy stored in the electromagnetic fields observed by an uniformly moving frame 
is related to that of the rest frame by  a factor different from $\gamma\,$. 
 Considering the energy and the stresses in the capacitor shells we show that the total energy of the system satisfies the expected relativistic transformation. 
 
\end{abstract}

\maketitle

\section{Introduction}
A very interesting example of an apparent paradox  in the relativistic transformation of the energy was given by Rindler and  Denur\cite{RD} some time ago. 
They considered a parallel-plate capacitor as seen from a frame moving in the direction of the electric field.
The electrostatic energy of the field configuration is reduced by a factor  $1/\gamma$ with respect to the
energy at the rest frame. This seems to contrast with the expected relativistic result of the energy increasing by $\gamma$. 
However, taking into account the stress in the braces that keep the plates at fixed positions they found
the appropriate relativistic transformation for the total energy of the system. 
This is a very nice example of the role played by the tension in the relativistic dynamics of extended bodies.

In the case of the parallel-plate capacitor, only electric fields show up in both frames and the geometry is unchanged by coordinate transformation.  When we consider a spherical capacitor, the change of observer has non-trivial effects in the geometry and in the field configuration. 
The spherical symmetry of the rest frame is not present for a moving observer. Furthermore, the field configuration for any moving frame will involve  magnetic fields in addition to electric fields.  Another important difference is that the relevant stress tensor contributions are distributed along the spherical capacitor shells while in the parallel-plate case they were acting just on the braces. 
We will see that the electromagnetic energy will transform with a factor different from $\gamma$ but also 
different from the parallel plate case. 
An interesting aspect that will emerge is that the tensions inside the capacitor shells are not 
uniquely determined. 
Only their boundary values are fixed by the electric forces on the charge distribution. 
However, the transformation of the total energy of the system depends only on this boundary values.

\section{Electromagnetic Energy }

\begin{figure}[htb]
\begin{center}
\includegraphics[width=0.5\textwidth]{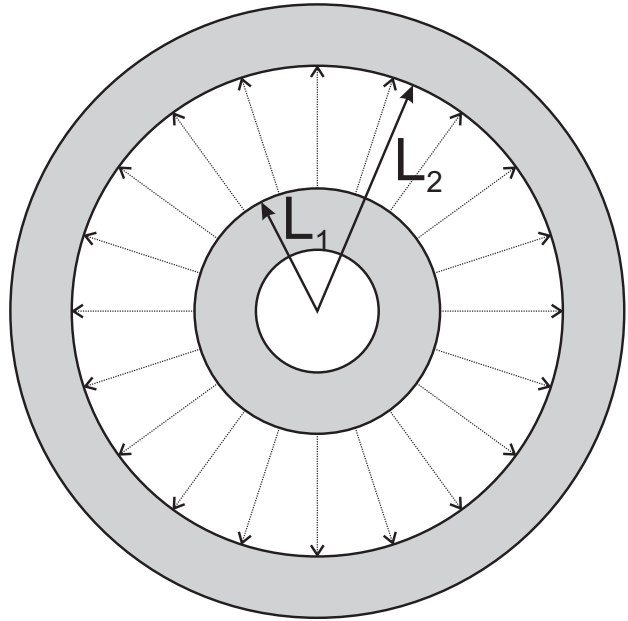}
\caption{Capacitor in the rest frame $S_0$. }
\label{CapacitorEmS0}
\end{center}
\end{figure}

\begin{figure}[htb]
\begin{center}
\includegraphics[width=0.5\textwidth]{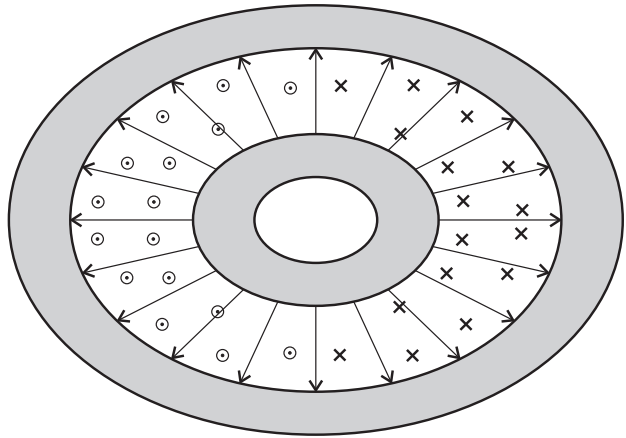}
\caption{Capacitor moving upwards in the frame S with electric and magnetic fields. The magnetic field lines are circles around the direction of motion.  }
\label{CapacitorEmS}
\end{center}
\end{figure}

Let us consider a spherical vacuum capacitor of internal radius $L_1$  and external radius $L_2$,
with charge  $q$ on the inside shell and $\,-\, q$ on the outside, as shown in figures \ref{CapacitorEmS0}
and \ref{CapacitorEmS}. 
$S_0\,$ is the rest frame of the capacitor and $S$ a frame moving with constant velocity $\,-\, v \, \hat z_0\,\,$ with respect to $S_0 \,$. 
The coordinate systems on the frames are parallel and the origins coincide with the center of the capacitor at $t_0 = t = 0$. While for $S_0$  the capacitor shells are spheres with center at coordinate origin, 
for  $S$, by Lorentz contraction, they are ellipsoids centered at $\,v\,t\, \hat z\,\,$ 
and with semi-axis  $L_1 (1,1,\gamma ^{-1})$ and  $L_2 (1,1,\gamma ^{-1})$ in the $(x,y,z) $ directions respectively.

For the $S_0$ observer the fields and the electromagnetic energy between the shells are just

\begin{eqnarray} 
\vec{E_0} &=& \frac{1}{4\pi \epsilon_0} \, \frac{ q \, \hat r }{ r^2 } \hskip 1cm , \hskip 1cm
\vec {B_0} \,\,=\,\, 0 \nonumber\\
U_{0\,_{EM}} &=& \frac{ q^2}{8\pi \epsilon _0 } \left(\frac 1 {L_1} - \frac 1 {L_2}\right)\,\,,
\end{eqnarray}

\noindent while for $S$ the fields, obtained from the relativistic transformations,
 are
\begin{eqnarray}
\vec{E} &=& \left(\, \frac{q\gamma }{4\pi \epsilon
            _0} \right)\,\, \frac{x\hat x + y\hat y + (z-vt)\hat z}{(x^2
            + y^2+\gamma ^2 (z-vt)^2)^{3/2}}\nonumber\\
& & \nonumber\\
\vec{B} &=&  \left( \,-\, \frac{q\gamma v }{4\pi \,c^2\, \epsilon_0} \right)
\,\, \frac{ \, \, y\hat x \,-x\hat y \,   }{(x^2 + y^2+\gamma ^2 (z-vt)^2)^{3/2}}\,.
\end{eqnarray}

The electromagnetic energy involves integrating over the region between shells. 
This can be done by introducing an ellipsoidal coordinate system:
$\, x \,=\, r\, \sin \theta \,\cos \phi \,,\,
y\,=\,  r\, \sin \theta \, \sin \phi \,,\,
z \, = \, \gamma ^{-1}r \cos \theta + v t \,$. 
The energy density in these coordinates is 

\begin{equation}
u_{_{EM}} \,= u_{_{EM}} (r, \theta\,)\,=\, 
\frac{\epsilon _0}{2} \left(\frac{q\gamma}{4\pi \epsilon _0 } \right)^2 r^{-4} 
(1 + \frac{v^2}{c^2} (\sin ^2 \theta - \cos ^2 \theta)\,)
\end{equation}

\noindent and the total energy is 

\begin{equation}
\label{EMEnergyS}
 U_{_{EM}} \,=\,  \int_0^{2\pi}\int_0^{\pi}\int_{L_1}^{L_2}\,\,
u_{_{EM}} (r,\theta\,)\, \gamma^{-1}\, r^2\sin \theta \, dr \,
            d\theta \, d\phi \,\,
=\,\, \left(\frac{q^2\,\gamma}{8\pi \epsilon _0 } \right)
            \left(\frac 1 {L_1} - \frac 1 {L_2}\right)
            \left( 1+\frac{1}{3} \frac{v^2}{c^2} \right)\,.
\end{equation}

So $\,\, U_{_{EM}} \,$ is not equal to $ \, \gamma\,  U_{0_{\,EM}}\,$.  Although the whole capacitor is an isolated system, the electromagnetic fields are interacting with the capacitor shells. 
Let us see what happens with the transformation of the energy of the shells.

\section{ Energy of the shells  }

For an extended distribution of matter, the energy density and the stresses  transform as components of a quadri-tensor of  rank two\cite{Rindler}. In particular, for a boost in the $z$ direction the matter density $\rho$ in the $S$ frame is related to the rest density $\rho_0$ by
 
\begin{equation}
\label{energianaoisolado}
c^2\rho =\gamma ^2(c^2\rho _0+  \frac{v^2}{c^2} T_0^{zz})\,\,,
\end{equation}

\noindent where $T_0^{zz}$ is the component of the stress in the $z$ direction, 
through a surface normal to $z$ in the rest frame $S_0\,$. The structure of stresses in this frame
can be described considering that the conducting shells are subject to electric forces that generate external tensions  on the outside surface of the internal shell and on the inside surface of the external shell. The spatial sector of the stress tensor in standard spherical coordinates has the general form

\begin{displaymath}
T^{ij}=
           \left(\begin{array}{ccc}
                T^{rr} & T^{r\theta} & T^{r\phi} \\
                T^{\theta r} & T^{\theta \theta} & T^{\theta \phi} \\
                T^{\phi r} & T^{\phi \theta} & T^{\phi \phi} \\
            \end{array}\right)\,\,,
        \end{displaymath}

\noindent where $T^{ij}=T^{ji}$. In the rest frame $S_0$ spherical symmetry implies that
$T^{r\theta}=T^{r\phi}=T^{\theta r}=T^{\theta\phi}=T^{\phi r}=T^{\phi \theta}=0$ and $T^{\theta \theta}=T^{\phi\phi}$ and that  all tensor components depend only on $r$. So,
        \begin{displaymath}
            T^{ij}\equiv 
            \left(\begin{array}{ccc}
                T^R (r)& 0 & 0 \\
                0 & T^L (r)  & 0 \\
                0 & 0 & T^L (r) \\
            \end{array}\right)\,\,.
        \end{displaymath}

\begin{figure}[htb]
            \begin{center}
            \includegraphics[width=0.6\textwidth]{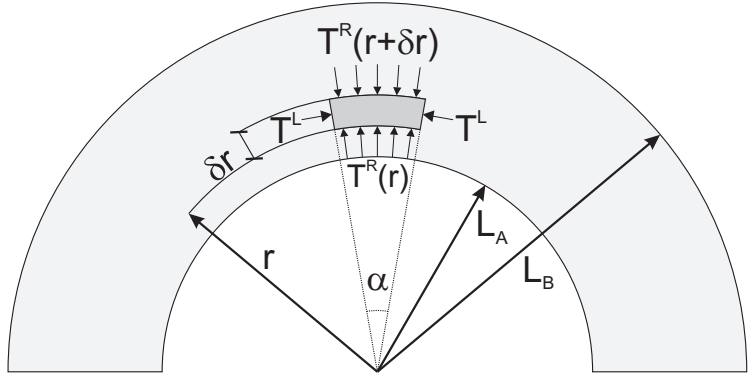}
            \caption{Forces acting on a volume element inside a shell.}
            \label{Stresses}
            \end{center}
        \end{figure}

The stresses are subject to the condition of equilibrium of forces in any small volume element inside a shell
as shown in figure \ref{Stresses}. The radial components of the forces acting on such a volume element are

\begin{eqnarray}
           F_2 &=& \int_0^{\alpha} (T^{R}(r)\cos \theta)(2\pi r\sin
         \theta)(r\mathrm d\theta)\,=\,  T^R(r)\pi r^2\sin^2\alpha\\
            F_1  &=&-  T_{R} (r + \delta r )\pi(r + \delta  r )^2\sin^2\alpha\\
            F_L &=& \int_r^{r + \delta r} (T^L(r^\prime)\sin
            \alpha)(2\pi)(r^\prime\sin \alpha)\mathrm dr^\prime \,=\,
             2\pi \sin^2 \alpha \int_r^{r+ \delta r} T^L(r^\prime)r^\prime \,  dr^\prime\,\,.
\end{eqnarray}

     The equilibrium condition  $F_1 + F_2 + F_L = 0$ leads to  
        \begin{equation}
        \label{eqtensoes}
        \frac d {dr}(T^R(r)r^2)\,=\, 2 \, r \, T^L(r)\,\,.
        \end{equation}
This equation is not sufficient for finding out the stresses inside the shells. 
However, as we will see,  it tells us that the total energy in the moving frame depends only 
on the boundary values of the stresses. The cartesian component, relevant for the total energy in equation
(\ref{energianaoisolado}) is
\begin{displaymath}
T^{zz}(r,\theta)= T^R (r) \cos^2 (\theta ) \,+ \, T^L (r) \sin^2 (\theta )\,\,.
\end{displaymath}

\noindent For each spherical shell the spatial integral of this stress, using equation (\ref{eqtensoes}) and
integrating by parts is 
\begin{eqnarray}
\label{casca}
U_{T} &=& \int_{L_{A}}^{L_{B}} \int_0^{\pi} \int_0^{2\pi} 
\gamma^2 \frac{v^2}{c^2} \Big( \, T^{zz}(r,\theta) \,   \Big) 
\gamma^{-1} r^2 sin\theta   d\theta d\phi dr \nonumber\\
&=& \gamma \frac{v^2}{c^2} \,\frac {4\pi}{3} \Big( (L_{B})^3 T^R(L_{B}) \,-\, (L_{A})^3 T^R(L_{A}) \,\Big)\,\,.
\end{eqnarray}

\noindent where $L_{A}\,$ and $L_{B}\,$ are the internal and external radius of the shell. 


In order to calculate the tensions on the surfaces of the shells we first have to determine the  
electric field acting on these surfaces. We scan start considering just a single spherical surface of radius $L$ with center at $r= 0$ and uniform charge density $\sigma\,=\, q /4 \pi L^2 \,$.
The electric field for points inside ($r<L$)  or outside ($r>L$)
this spherical surface can be calculated using directly Gauss law. One finds the well known result 
\begin{eqnarray}
\vec E (r) &=&  0\,\,\hskip 3cm (r < L)\nonumber \\
\vec E (r) &=& \frac q {4\pi\epsilon_0 
 r^2}\hat r \, \hskip 2cm (r > L) .
\end{eqnarray}

For points that lie on the surface:  $\,r = L\,$, we can calculate the electric 
field by consi-dering an infinitesimal element of area $dA$ and integrating the electric field 
contributions $d \vec E$ produced by all the other area elements of the surface. 
The result of this integration, in the limit where the area $dA$ goes to zero  is a total electric field
reduced by a factor of 1/2 with respect to the field just outside the surface:

\begin{equation}
\label{fieldonshell}
\vec E = \left(\frac 1 2 \right)\frac q {4\pi\epsilon_0 
L^2}\hat r\,\, \hskip 2cm (r = L)\,\,.
\end{equation}

\noindent Now, returning to our system of two shells, the external tensions act only
on the surfaces where the charges are located. That means: the
external surface of the internal shell and the internal surface of the external shell.
So we will consider only these surfaces. 
For the inside shell there is no contribution from the 
external shell, so the field on the charged surface is just 
\begin{equation}
\vec E (L_1)= \left(\frac 1 2 \right)\frac q {4\pi\epsilon_0
L_1^2}\hat r \,\,. 
\end{equation}

\noindent While for the external shell we have the superposition of the field produced by the internal shell (positive radial direction and no factor of 1/2) and the field produded by the
external shell (negative radial direction and factor 1/2) resulting in:

\begin{equation}
\vec E (L_2)= \left(\frac 1 2 \right)\frac q {4\pi\epsilon_0
 L_2^2}\hat r \,.
\end{equation}

 The radial tensions on the surfaces of the shells correspond to the inward radial components of the 
forces per unit of area. So, for each surface we multiply the electric field by the surface charge density
and by the unitary normal pointing inside the shell.
We find that the nonvanishing tensions are equal to
 
\begin{equation}
 T^R (L_1 )\,=\,  -\frac{q^2}{32\pi^2 L_1^4\epsilon_0}
\,\,\,\,; \hskip 2cm
T^R (L_2) \,= \, - \frac{q^2}{32\pi^2 L_2^4\epsilon_0}\,.
\end{equation}

So, integrating the energy density (\ref{energianaoisolado}) in the volume of the two shells, we find contributions to 
eq. (\ref{casca}) from the internal shell with $L_{B} = L_1 \,$ and from the external shell 
with  $L_{A} = L_2 \,$
The energy of the shells in the $S$ frame is thus 

\begin{equation} 
\label{shellenergy}
M c^2 \,=\, \gamma M_0 c^2 \,-\,\frac{1}{3}\, \frac{v^2}{ c^2} \,\,  \frac{q^2\gamma}{ 8 \pi\epsilon_0}
            \left(\frac 1 {L_1} - \frac 1 {L_2}\right)\,.
\end{equation}

\section{Total Energy}

The total energy of the capacitor (shells and fields) using eqs.   (\ref{EMEnergyS}) and (\ref{shellenergy}) 
is   
        \begin{displaymath}
            U\, = \, U_{_{EM}}\, + M c^2 \,
                     = \gamma\, \Big( M_0 c^2 \,+\, U_{0_{\,EM}}\, \Big) \,=\, 
           \gamma \, U_0\,.
        \end{displaymath}

So, the total energy transforms in the expected way. Also, it does not depend on the stresses inside the capacitor shells but only on their boundary values.  This fact is due to the asymmetric time delay in the  electromagnetic forces acting on the different parts of the shells in the $S$  frame when the capacitor is charged, which leads to a net work in this frame.

\acknowledgements The authors are partially supported by CNPq and FAPERJ.

\end{document}